# A FRAMEWORK FOR SECURE DIGITAL ADMINISTRATION[1]

## Diana Berbecaru    Antonio Lioy    Marius Marian

*Dip. di Automatica e Informatica, Politecnico di Torino, Torino, ITALY*
*e-mail: {berbecar, lioy}@polito.it, marian@athena.polito.it*

## Abstract

The efficiency and service quality in public administration can be improved by using electronic documents (or e-docs) and digital signature to speed up their activity and at the same time to better satisfy customer needs. This paper presents an XML-based document exchange system that integrates a platform for the management (creation, search, storage) of e-docs and a secure trustworthy environment for digitally signing e-docs. The framework provides security services, like privacy, authorization, authentication and non-repudiation. Possibly a mobile terminal equipped with a smart-card reader and integrating WYSIWYS features could be used for viewing and signing e-docs. The proposed system can be easily integrated with the infrastructure (e.g. database system) already in use at each administration site. It is described also the use of the system in a real world service.

# INTRODUCTION

The growth of IT technology and especially of Internet has created the opportunity to improve businesses' and citizens' access to information and regulation and to facilitate contacts, exchanges and feedback between administrations and third parties. However, announced and expected change of the delivery, availability of information, transactions and services are shaded by the lack of appropriate standard formats for data and signatures to be used, concerns about security and legal uncertainty.

Digital signatures can provide authentication and integrity functions, using signature techniques based usually on asymmetric signature keys certified by a trusted Certification Authority (CA). Improved formats of digital signatures (called electronic signatures, or ES) have begun to be defined since the European Parliament and the Council of European Union established the Directive on a common framework for ES [EUdirective]. Probably ES laws in different countries will lay also the foundation for the issuing of e-docs by public institutions (e.g. administrative bodies, universities) which - digitally signed - will be held equal to conventional paper-documents. Nevertheless, so far a standardized format to be *specifically* used for e-docs hasn't been defined.

---

[1] This wok has been funded by the European Commission under project AIDA (IST-1999-10497).

Complexities are met both in the new field of ES and in its applications and during the actual design and functional decomposition of a system providing for intelligent e-administration. Integrated systems must satisfy the requirements related to the generation of e-docs, display of e-docs in a way that makes signing of such e-docs free from unpleasant surprises and construct enhanced formats of digital signatures with different legal values. Such systems must be easily incorporated into already existent administration body' infrastructures (e.g. a database system). Moreover, with the increasing number of persons using mobile devices (e.g. personal digital assistants – PDAs), the challenge is to improve the functionality of these devices by allowing their users to securely view and sign e-docs. The Advanced Interactive Digital Administration (AIDA) system, developed inside the homonymous EC-funded project [AIDA], takes up the task of creating a framework with the features described above. It implements a secure technical environment for e-administration and uses it to demonstrate the feasibility of trustworthy digital signatures in combination with enhanced e-docs.

## GENERAL FRAMEWORK REQUIREMENTS

The traditional public administration processes are based on paper documents as illustrated in Figure 1. It can be noticed that the main entities generally involved are the citizens and the companies (or end-users) and the authority (or administration). The paper-based systems usually present several disadvantages due to large amounts of papers handled, delays in documents delivery and because the interface between the end-users and the authority is primarily done on person or via mail. An electronic system has to support public administration services by using e-docs, digital signatures, open networks like Internet, electronic registries and other modern tools. The system should provide services for both e-administrators (i.e. persons that administer the electronic platform) and for the users and should be usable by any e-administration body, no matter if it is a university, a public body or a company.

The basic design requirement of such a system is *flexibility* to help the e-administrators in easy configuration and management and the users in easy but secure e-doc creation and signing. Another primary requirement is *modularity*. This means that all management solutions have to maintain the services offered while incorporating new versions and variations. Finally a fundamental consideration is to implement the framework to meet an adequate level of *security*, which is an essential property for the system to be effective. Among the security requirements we can identify: authentication, authorization, integrity, confidentiality, data archiving, chain of custody, non-repudiation. Paper documents pass through pre-defined hierarchies that have to be maintained also for e-docs. The term *chain of custody* refers to these rules that are typically enforced using application logic and not cryptography. Authentication must be done both for the end users and for e-administrators. When accessing a service a user needs to authenticate himself to the E-administration

Service Provider (ESP) and in some cases the ESP has to authenticate also to the users. Confidentiality is ensured by the encryption of data between clients and servers to prevent its exposure over public Internet links.

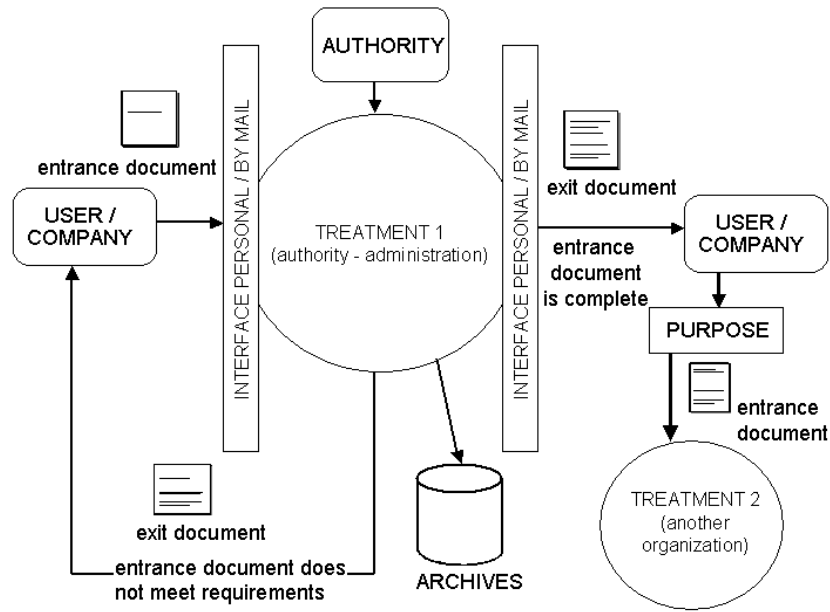

**Figure 1** – General paper model in public administration

The framework is composed of several services to achieve the design goals. The e-administrators require services like e-doc creation, revocation, e-doc repository and configuration services as well as security service. Users are provided instead with e-doc creation, security, signature and e-doc validation services. *E-doc creation* is performed either via editing already created e-docs or are generated by a system component from data provided by the user and/or extracted from databases. The *signature service* permits users to digitally sign e-docs by using a trustworthy environment for the creation/verification of digital signatures. The term *digital signing* designates the process by which a digital signature can be affixed to an e-doc. *E-doc revocation service* refers to the possibility to revoke e-docs, the *repository service* ensures the availability of structures used for the creation of well-formed e-docs put to users' disposal, as well as storage of signed e-docs in a directory. The *e-doc validation service* is used to ensure that an e-doc is valid at a certain time. The users are guaranteed that an e-doc has exactly the content specified by an e-administrator in an appropriate structure. For example no hidden fields are allowed in an e-doc. The same service has to provide also techniques for long-term validity of signed e-docs like signing the complete signed document with a stronger algorithm before the old algorithms cease being secure. The e-administrators can set up the platform using the *configuration service*. The *acknowledgement service* is providing a signed feedback to the user confirming the receipt of the e-doc. On the other side, in the same service, it is necessary to supply the means for logging the process' progress (log module). *Security service* is required both by users and e-administrators. It has to ensure maximum security for access and information management and must respond to the security requirements mentioned above. *Publishing service* allows each e-administration body to freely use a directory most suitable for their needs to make information public.

## AIDA OVERVIEW

Many pieces already exist, various document management systems implement organizational workflows in proprietary manner. To enable AIDA a number of tasks have to be addressed. First is a trustworthy environment must be defined and implemented. Another task is to define machine-readable data structures for e-docs. In the end, a management platform for e-docs has to be implemented, that is an environment for generation of e-docs, including a secure infrastructure that support the whole life cycle of an e-doc and a platform utilized by users to display and verify the e-docs' content and signatures applied on them. The system must use the back-end and front-end already existent and in use at each administration side. This means for example that the public bodies don't have to modify their currently used Database Management System when adding support for AIDA. To create a trustworthy environment, all components have to be trustworthy to contribute to an overall secure solution. The signature creation device (SCD) must make forging of signatures as difficult as possible by requiring an active user confirmation before signing anything. Entering a PIN, a fingerprint or any other authenticating information should ideally be done directly on the SCD. The SCD must be embedded into a secure environment that also makes forging signatures difficult. This environment is responsible for preparing the data to be signed and to be sent to the SCD for hashing and signing. As the SCD might not be capable of displaying the data to be signed, the security of this environment and the binding to the SCD is crucial for the overall functionality of the system. [Heiberg2000] explains clearly the types of attacks that can be performed on signing environments. Ideally, the SCD and the signature environment are integrated into one component. Displaying the data correctly is also very important. *Correctly* means that the user sees exactly the data he is going to sign and nothing else. A secure signature environment must therefore be able to parse and understand data and reject signing data that contain illegal parts, which are unknown to the display unit. Examples of illegal contents are white text on white background or unknown HTML-tags etc. We call this module the What You See Is What You Sign (WYSIWYS) module.

AIDA's objectives are to create and to incorporate all existing components needed to provide secure signature environment. It will provide a software-only solution and an appropriate solution for secure signing equipment. For example, the software-only solution is to be used with different platforms and could differ from the solution for secure signing equipment which could be limited by power, computational or network capabilities of the signing device used. To a large extent, a software-only solution must rely on the environment, as it contributes largely to the security of the system. Even if such a solution cannot provide perfect security sometimes this is not absolutely required. Features like the ability to display the data to be signed accurately will still be very useful and appropriate for mass-market. Alternatively, small personal devices, like notebooks or PDAs that are equipped with a smart-card reader and that integrate WYSIWYS features, seem to be the ideal solution for secure signing equipment: the devices are small and can be carried around easily, while the display is usually still capable of displaying different forms of data to be signed [Heiberg2000, Boneh1999].

## PRESENTATION OF THE AIDA ARCHITECTURE

AIDA framework conforms to a web application architecture, which typically follows an *n*-tier model (Figure 2). The first tier is the presentation layer, which includes not only the web browser but also the web server, which is responsible for assembling the data into a presentable format. The second tier is composed of the application layer and consists basically of the code, which the user calls upon through the presentation layer to retrieve the desired data. It is further the task of the presentation logic to receive the data and format it for display. The third tier contains the data that is needed for the application and could consist of any source of information. The data source is not limited to just an enterprise database like Oracle but could include also a set of XML [Bray1998] documents or a directory service like an LDAP server. The middle tier is extended to allow multiple application objects. The application objects each have an interface, which allows them to work together. In our particular case, the database backend already existent at an ESP will work together with AIDA server (or AIDA platform, named further *Aplatform*) through interfaces implemented at application layer to supply data to the presentation layer. The presentation and the application levels are incorporated into the *Scenario Application Framework* (SAF) as illustrated in Figure 3.

*Functionality*. Client-server paradigm is used to implement the overall system. This architecture also takes the form of a server-to-server configuration that is one server plays the role of a client and request services (e.g. database service) from another server. The *Aplatform* communicates via AIDA commands (or *Acommands*) with client applications using a nonstandard application protocol named *AIDA protocol* (or *Aprotocol)*. Every *Aplatform* operation is explicitly requested by the SAF, which implements the actual service workflow. The SAF is responsible for interaction with the web-server component that will surely be present in any scenario to provide the user with an appropriate web portal. It gathers also proprietary data through appropriate interfaces and communicates with the *Aplatform* for specific functions like e-doc creation, display and verification. Two equivalent application ports with similar functionality, called the *scenario* and *service ports*, are used to access the *Aplatform*. These ports can be configured to allow an ESP to freely choose the openness of the system that is they are used to receive or send back messages directly or via gateway programs. The ports can be set up for two types of settings: to specify the set (full or restricted) of *Acommands* accepted by the port and to indicate the outside visibility of the port. The gateway programs are simple programs that tunnel commands from one port to another one and connects clients situated outside the ESP's intranet to the *Aplatform*. This permits the SAF to be situated anywhere in the Internet or in the ESP' intranet. The *administration port* is a special port used to start/stop the other ports and to accept *Aplatform* administration functions.

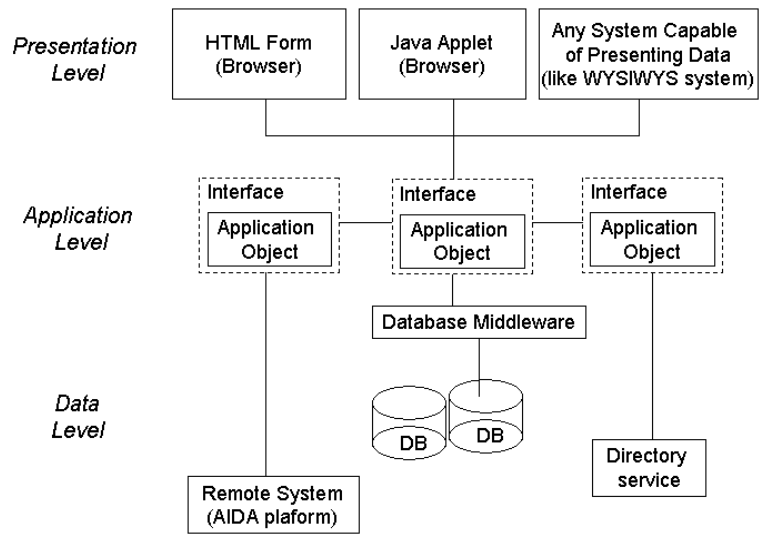

**Figure 2** – General architecture

The *Aprotocol* uses signed XML structures for the AIDA messages (commands, responses) exchanged. A special Document Type Definition (DTD) is used for the syntax of these messages. AIDA messages can be sent directly to the *Aplatform* using the *Aprotocol* or tunneled via HTTP to the gateways that send them to the *Aplatform* on dedicated ports. The Security Service Bricks (SSB) are the modules that support the security service and enable most of AIDA' secure functionality like support for smart cards, SSL or XML signatures.

*Smart cards.* Smart cards are secure, portable computing devices capable of storing sensible information and compute digital signatures and other cryptographic primitives [Vedder1998, Okamoto1999]. This project considers the use of smart cards for two purposes, that is cards are used to uniquely identify a user inside an organization (authentication) and to produce digital signatures by storing private keys and the corresponding X.509v3 public key certificates.

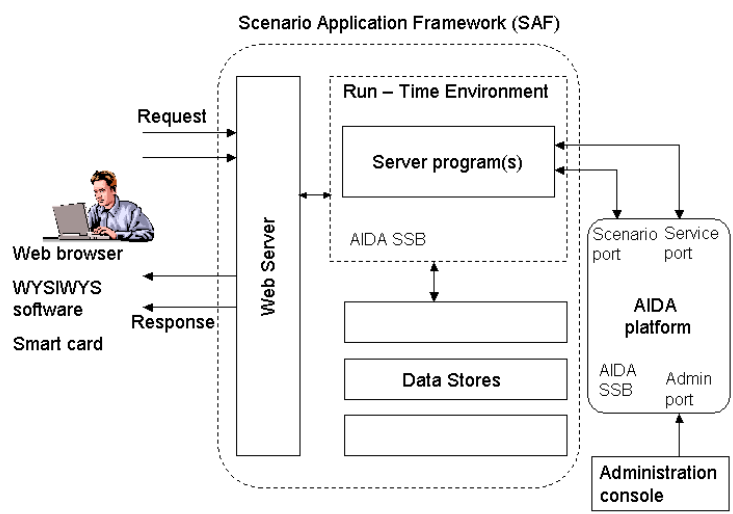

**Figure 3** – Interconnection between Scenario Application Framework and AIDA platform

*Authentication.* AIDA employs the Secure Socket Layer [SSL] to ensure channel integrity, authentication and privacy. Server and client authentication are based on X.509v3 certificates. Server authentication enables clients to verify the server they are communicating with. Client authentication allows servers to verify the client's identity and use it as a basis for access-control decisions. One aspect to keep in mind is the separation of keys for authentication purposes and for digital signing when the system has to be conformant with signature laws. Thus, different key pairs must be used for those purposes and the PINs to trigger the creation of a signature must be different for these keys. One difficulty that has to be addressed at the level of the SAF is the user's authentication that is verifying the identity of the client interacting with the *Aplatform*. For this purpose the user's certificate has to contain public data uniquely identifying him in an organization, like an identity card number. To avoid the insertion of the ID data in the user's certificate (because for example some users simply don't want their identity card number to be public) alternative solutions would be to use an explicitly map a user's certificate to user ID data in specialized store on the *Aplatform* or to put user ID data on the user' smart card into special files instead inside the user's certificate. This is called a *personalized* smart card because it contains files with identification information chosen by the institution. The SAF checks the user's identity. Consequently, it is up to ESP to personalize the smart cards such that the SAF can easily use the data on the card.

*Role system*. Every command sent to the *Aplatform* is signed with the private key paired to a public key certificate used for authorization purposes called *role certificate*. These can be any certificates the ESP wants to use and it is not compulsory for them to have high level of assurance or to be issued by certain CAs. The *Aplatform* checks whether the *role* has the privileges to execute the requested command by looking at the *role map*. This *map* assigns each role with all *Acommands* and e-doc types that are allowed for it. Most of the commands have also e-docs that they operate on. So, besides allowing only certain commands to be executed by a specific role, the *Acommand*'s execution can be further restricted by defining a set of e-doc types it is allowed to operate on. This mechanism allows different users to use the same *Acommand* set but each one can only work on the specific type of e-docs he is responsible for.

*E-Doc Format*. The system is not intended to introduce completely new and proprietary document formats: AIDA uses XML [Bray1998] for several reasons. First, this format supports strict separation of content and presentation. Document formats in use today (PDF, Word, HTML) usually mix up data and presentation, making them inappropriate to use when the same information need to be presented differently. Any format mixing presentation and content and that allows active content is also risky to sign: macros, white fonts on white background are only some of the threats. Moreover, some of the formats (e.g. Word's .doc files) have non-public definitions and hence cannot be used in security-critical applications. Another reason is that XML allows also including the DTD into the e-docs. Thus, attacks as described in [Heiberg2000] performed by a malicious individual when using HTML documents, will be avoided. The attack is based on the fact that an HTML document is viewed differently with an ASCII-editor as a source code and with a web browser and an attacker could include text as comments. Web browsers do not render comments and signer cannot view what he's signing unless he views the source file. When using this format, a display module can apply advanced filtering techniques for displaying XML documents (or XML docs) in a trustworthy manner to be used for signing and verification [KarlS2001]. Consequently, AIDA is focused on XML docs and on signed XML docs as well. One task to be solved when using XML for secure information exchange is the implementation of a standard for signed XML as defined in [Eastlake2001] and this is incorporated into the WYSIWYS module. Another task is the conversion of relational database data to and from XML in the simplest way.

## AIDA Platform

The *Aplatform* (Figure 4) is located inside the ESP's intranet and in real-life scenario it should run on a dedicated machine in a high-security area. The server is installed and configured using the administration console by the e-administrator, under strict security measures. Example configuration tasks are setting and updating the role map or specifying the data types and styles for the e-docs. The ESP applications, generally called *Scenario Applications* (SA), implement the workflows for various e-doc services. For this purpose the SAs interact with the *Aplatform* and with the user interfaces (browser, WYSIWYS client). The *Aplatform* must not be confused with the web server that interacts with the users and runs SAs that in turn communicate with the inside *Aplatform*. The Application Server (AS) acting as a web server is the connection point for service users while the *Aplatform* is never visible externally. The d*efinitions repository* contains specifications and meta-information for the creation and verification of XML docs, like DTDs and stylesheets and internal data for the authority. The *document directory* contains every signed e-doc created by a user or organization.

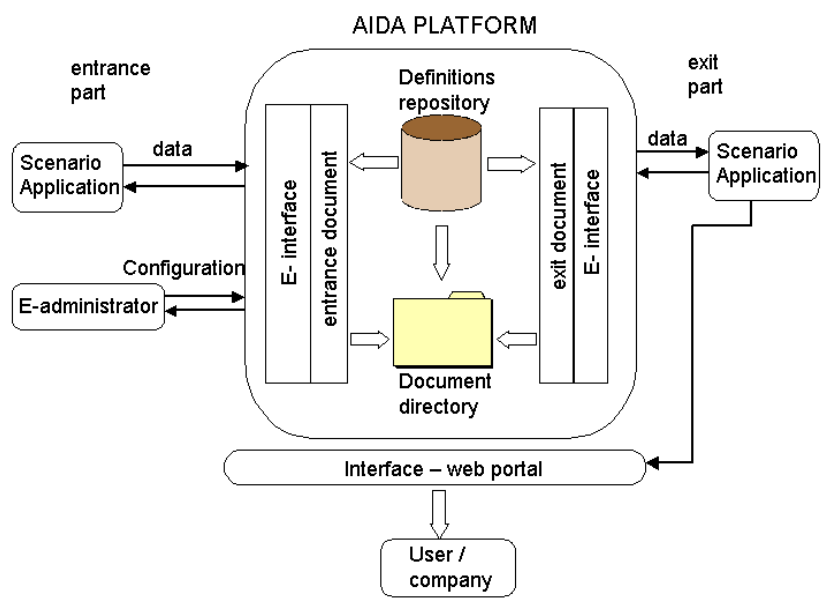

**Figure 4** – AIDA platform

# WYSIWYS Module

The AIDA's WYSIWYS software (client) satisfies some rough requirements:

a)  View e-docs. The device should display the e-docs in a reliable and trustworthy fashion so that the user can view it.
b)  Sign e-doc. The signature should be authentic and not forgeable.
c)  Verify e-docs. The system should be able to verify a signed document and display the result in a clear and unambiguous way.
d)  Create legally binding signatures. Official parties should accept signatures created with this system.
e)  Little or no *maintenance required*. The program should not require regular program update or installation or if absolutely necessary this should be as easy as possible.

The modularity of the WYSIWYS code reflects the system structure in a clear way. Even though the system is primarily focused on XML docs, it should be possible to extend it to support other document formats as well. The WYSIWYS module incorporates a XML viewer for displaying XML docs, a library for signing XML docs and a cryptographic library for digital signing that makes use of smart cards to generate signatures. A solution for enabling Java applications to create digital signatures directly on cryptographic smart cards is shown in [KarlS2000]. The framework consists of several packages including only interfaces, abstract classes and generic code common to all documents. Classes that are needed for all documents are placed in separate packages. Examples of such classes are classes to access services of the *Aplatform* like getting the e-doc' status.

## Scenario Desktop

The Scenario Desktop (SD) is a Graphical User Interface (GUI) tool for interactive operations and handling of e-docs. Persons called *referents* that are responsible for certain tasks use this tool to model the operations in the ESP's workflows that need human processing. The referents can view, create, store, sign and search for specific e-docs stored on the *Aplatform*. To modify e-doc' status the e-doc *attributes* are defined. They allow the *Aplatform* to fulfill a variety of functions and support various models of e-doc structures, meta-information and e-doc workflows of various ESPs. No single system could be built to incorporate every aspect of e-doc management preferred by an ESP and his existent business logic. To handle *Aplatform*'s data stores as needed (e.g. to partition the document directory into *input* and *output* e-doc parts), to add specific data to e-docs (e.g *state*) or to configure type-specific data need by the SAF (e.g. to mark some fields as optional) the so-called *attributes* are used to hold information needed by the ESP. AIDA uses three levels of attributes whose description and usage is not part of the present paper. To understand the case study presented further, we'll just state that an attribute is used to store the e-doc' status (e.g. *pending*). Unlike *static* attributes, which are set once at the *Aplatform* start-up and never changed later, these are *dynamic* attributes because they can be read and written by the SD. Referents can create *output* e-docs using field data extracted from one or more existing XML e-docs serving as input source called *input* e-docs as well as field data entered directly via the GUI. The connection between the *input* and *output* e-docs is not represented in the definitions repository but it is specified using *processing rules* inserted into an XML configuration file named *properties file*. Fields in the *output* e-doc not automatically filled by the application after applying the *processing rules* must be presented to the referent in a dialog box and their values must be entered manually.

The role system sets up restrictions on the e-doc types each referent is allowed to generate and view. The role certificates allow access control by restricting the referent's actions to the ones her employer (ESP) sets up according to his internal responsibility scheme. In practice, the SD signs every command sent to the *Aplatform* with a private key and attach the role certificate (containing the pair public key) to be checked against the *role map*. If the requested command and/or e-doc type is not allowed for that role, the request will be rejected. The SD interacts with the *Aplatform* either directly to the *scenario port* or indirectly via gateway programs that forwards requests to the *scenario port*. The latter case is very useful and necessary when the SD is run outside ESP's intranet. The communication between the SD (acting as client) and the *Aplatform* is completely transparent because it is handled within the application support library. This library provides a number of classes and methods that resemble functionality which is either directly executed on the client' side or implicitly involves an *Aplatform* communication using the *Aprotocol*.

# CASE STUDY

The implementation of a real service based on AIDA involves the interconnection of several components, some of which are not (and cannot be) part of the *Aplatform* itself. This holds especially for the actual workflow logic that connects to proprietary resources to query user data or other information necessary to complete e-doc processing. This section describes a pilot service based on AIDA to be used inside Politecnico di Torino (Polito), i.e. the exam admission service (EAS). This is a widely and frequently used service inside any education institute and is a good candidate for e-administration due to the large spectrum of users and paper documents usually involved.

## Current Service Organization

Inside Polito the requests for exam admissions are collected at the Student Service Office (SSO). The requests are made from a special networked terminal called *totem*. Totems are secure devices that act as front-ends for the SSO services and are similar in some way with the Automated Teller Machines (ATM) because they permit users to make payments too. Totems have dedicated communication links with servers inside SSO and with servers inside banks because some of the services available require payments. One of the services available is the EAS (Figure 5). After authenticating the student a number of checks completely transparent to the user are performed by the AS: the student's enrollment status, the rights to do the exam, the payment situation. Upon completion of these checks the totem generates an auto-sticking paper exam admission card (EAC) containing student and exam specific data. The EAC is valid for a whole exam session (i.e. 6 weeks, that includes 3 exam rounds) but getting the EAC is not a method for actually booking for a specific exam at a date with a certain professor. Besides the data regarding the student (e.g. ID number, place of birth), the faculty (e.g. name), the exam (e.g. and name, date), EAC related data (e.g. validity period), the EAC contains also two exam evaluation tickets (EETs). The EAC is signed by the responsible of the SSO. The EETs contain the same data for student, faculty and validity period as the EAC. Additionally they have blank fields to be filled in by the professor after the exam (e.g. the mark). At exam time each student must give the professor the paper EAC to prove that he is can do the exam in that session. After examination the professor fills in the blank fields in the auto-sticking EETs, puts them in the registry, writes some extra data in the registry (like questions put at the exam), signs the registry and finally sends it to the SSO to be processed. Processing the registry means tearing off the right part containing one EET and the extra data. Next the registry containing only one EET in the first column is returned to the professor for his archive and to use the remaining empty pages.

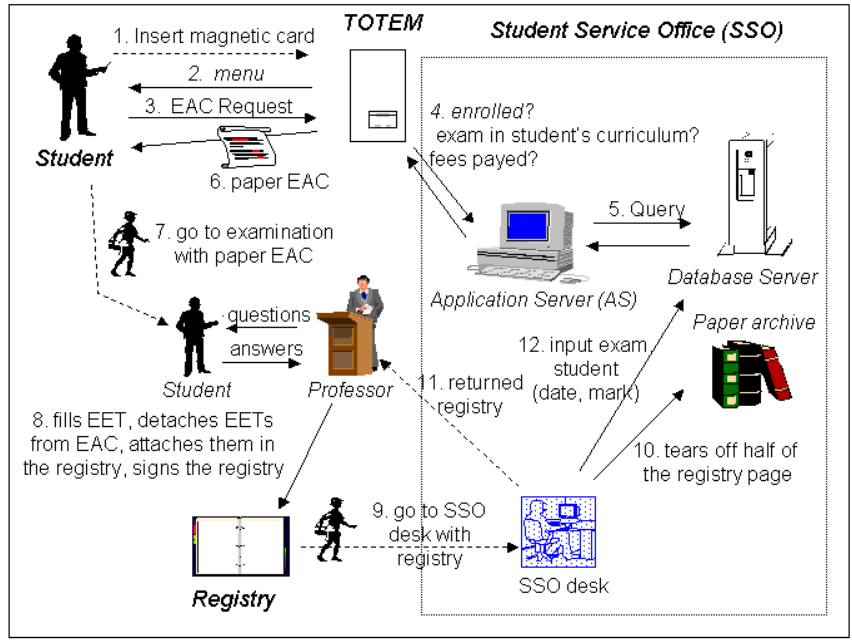

**Figure 5** – Current paper-based Exam Admission Service

# Proposed Service Organization

An improved service could be offered by using an appropriate signed e-doc instead of the paper EAC to provide proof of identity and authority (Figure 6). The student can continue to make exam admission requests using the *totem* and the web interface running on the client PC within the *totem*. Additionally the student can establish SSL-client authenticated sessions when connecting to the AS remotely (e.g. from home). Instead of the paper EAC the AS issues a signed exam admission card in electronic form (e-EAC). This e-doc contain only exam related data and to uniquely identify the student inside Polito and doesn't include any more data from the EET, which will be put in a separate e-doc called e-EET. At exam time the student can prove that he registered for the exam in two ways. Either he brings the e-EAC stored on a floppy disk and lets the professor check its correctness using a terminal (PC or mobile device) that runs WYSIWYS client. Another way would be for the professor to use the terminal to connect to the AS and extract the e-EACs via a *search Acommand* from the *Aplatform*, by using data provided by the student.

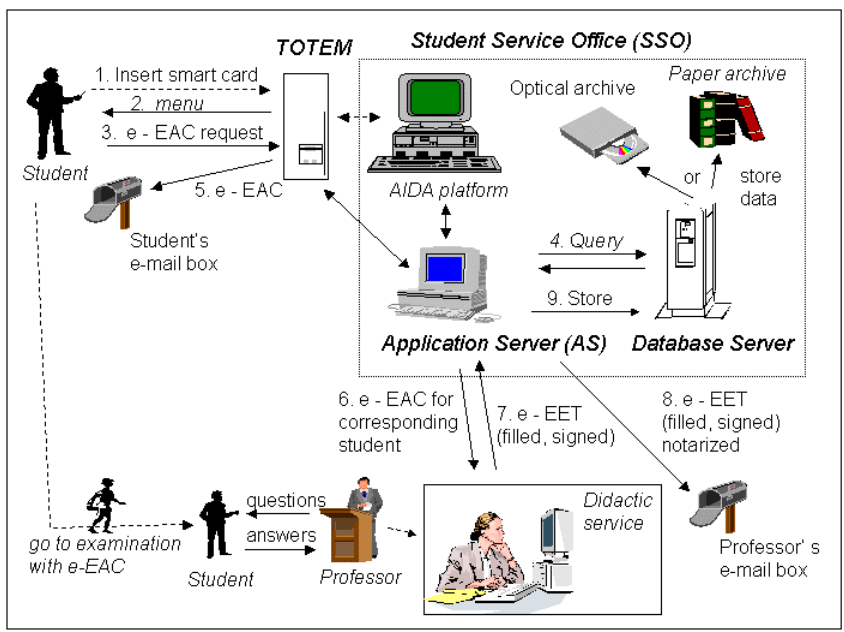

**Figure 6** – Proposed electronic Exam Admission Service

After the examination, the professor uses the SD to process all the e-EACs stored on the *Aplatform* for that exam. The e-EAC status of is kept in an *attribute* field. Normally all e-EACs for an exam should be *pending*. As result of the *search* command, one line is inserted in the working screen for every e-EAC to be processed, together with visible student ID data. The SD applies *processing rules* for generating an e-EET from an e-EAC. To accomplish this the SD calls first a support library function to open an e-EAC into separate fields. Inside the window screen there will be noticed field data extracted from e-EAC (as *read-only*) and additionally there are text boxes for the fields related to the exam (e.g. date, mark) to be still filled in by the professor. Upon data's insertion the SD triggers the WYSIWYS module that gets appropriate transformation rules for e-EET from the *Aplatform* according to the user/display capabilities set on professor's PC. At this point the professor can view and sign the e-EET. Finally the WYSIWYS client sends an answer in the form of signed e-EET (or error code) to the *Aplatform* to store it and to set its status to *issued*. The originally selected input e-EAC e-doc' status is set to *processed*. When using a mobile terminal the performed steps should be similar as described above except that the device's browser and an internal smart card reader are used. In the end the professor could get a copy of the signed e-EET sent by AS to his e-mail box and in the same time, another copy may be sent to a printer or to a hard disk for archiving.

# Implementation Aspects

The model adopted for Polito's infrastructure is Oracle's Internet Computing Architecture. It comprises a three-tier architecture with a thin client, an application server and an Oracle database on the back end. The client runs a Java application using a Java-enabled web browser and sends user requests to the form servers situated at application tier level. The client tier always talks to the application tier, never directly to the database tier. A totem is composed of a PC, a printer, an UPS, an adjustable monitor and is network connected. The Netscape browser is started in *super kiosk* mode allowing it to run in a frameless window. The SA is implemented as a set of Java servlets (gateway and application) running on the AS. They communicate on one side with the *Aplatform* and with the database server and on the other side with the user as a typical web application. This means that it collects data from the user via HTML forms, sends a request to the web server, runs the requested servlet, formats the output in HTML or in the format accepted by the AIDA's display unit and sends it back to the browser or to WYSIWYS module for display. Oracle Internet Application Server has been used for deploying and managing Internet applications inside Polito. The incoming requests are processed by Oracle HTTP Server (based on Apache) and are directed toward the Apache Jserv servlet engine via the Apache JServ protocol. Proprietary data are extracted from an Oracle database over the network using JDBC. With support for AIDA the present three-tiers architecture is extended to an n-architecture because the number of data sources increases. The web server performs client authentication to grant user's access to certain important functions for the workflow. Hence, issues like professor's identification and preventing one professor to see the examination data of another professor lies solely under the control of the SA. *Gateway servlets* tunnel *Acommands* exchanged with the *Aplatform* coming from outside the Polito intranet. When exam admission requests are made from home, the user must download and install the WYSIWYS software from the Polito's web site. Installation involves a registration process for automatic application launch from within the web browser on receipt of special file types. X.509 certificates are issued by the EuroPKI CA [EuroPKI]. This is a non-profit organization established to create and develop a pan-european public-key infrastructure and it is derived from the EC-funded ICE-TEL and ICE-CAR projects. It provides public-key certification services to the wide European Internet community to support borderless network security.

# CONCLUSIONS AND FURTHER WORK

This paper introduced the goals and the features of a general framework for e-administration and presented the design and work done so far in the AIDA project. The main objective of this project is to design a secure and flexible framework that enhances public services. The adopted solution is based on the use of e-docs in XML and embedded digital signatures and satisfies the main requirements of a secure administration system. The paper shows also how the outcome of the project can be applied to implement a complex case study, making it appropriate to any real social environment. Further efforts include extension of the system to support parallel signatures, chain of custody, enhanced formats of electronic signatures and mobile devices with reduced computing capabilities.